\documentclass[pre,twocolumn,superscriptaddress,showpacs,showkeys,floatfix]{revtex4}
\usepackage{listings}
\usepackage{fancyvrb}
\usepackage{graphicx}
\usepackage{latexsym}
\usepackage{bm}
\usepackage{natbib}
\usepackage{dcolumn}
\usepackage{amsmath}
\usepackage{amstext}
\usepackage[english]{babel}
\usepackage{flafter}
\usepackage{url}
\usepackage{multirow}
\usepackage{float}
\begin{document}

\title{Consensus clustering in complex networks}
\author{Andrea Lancichinetti}
\affiliation{Complex Networks and Systems Lagrange Lab, Institute for
  Scientific Interchange, Torino, Italy}
\affiliation{Physics Department, Politecnico di Torino,
Corso Duca degli Abruzzi 24, 10129 Torino, Italy}
\author{Santo Fortunato}
\affiliation{Department of Biomedical Engineering and Computational
  Science, School of Science, Aalto University, P.O. Box 12200, FI-00076, Espoo,
  Finland}
\affiliation{Complex Networks and Systems Lagrange Lab, Institute for
  Scientific Interchange, Torino, Italy}

\begin{abstract}
The community structure of complex networks reveals both their organization
and hidden relationships among their constituents. Most
community detection methods currently available are not deterministic,
and their results typically depend on the specific random seeds,
initial conditions and tie-break rules adopted for their execution. Consensus clustering
is used in data analysis to generate stable
results out of a set of partitions delivered by stochastic
methods. Here we show that consensus clustering can be combined with any
existing method in a self-consistent way, enhancing considerably both the stability
and the accuracy of the resulting partitions. This framework is also
particularly suitable to monitor the evolution of community structure in
temporal networks. An application of consensus clustering to a large citation network of
physics papers demonstrates its capability to keep track of the birth, death and diversification of topics.
\end{abstract}



\maketitle

\section{Introduction}
\label{intro}

Network systems~\cite{albert02,dorogovtsev01,newman03,pastor04,boccaletti06,caldarelli07,barrat08,cohen10}
typically display a modular organization, reflecting the existence of
special affinities among vertices in the same module, which may be a consequence of their
having similar features or the same roles in the network. Such
affinities are revealed by a considerably larger density of edges
within modules than between modules. 
This property is called community structure or graph clustering~\cite{girvan02, schaeffer07,
  porter09,fortunato10,newman12}: detecting the modules (also called
{\it clusters} or {\it communities}) may uncover similarity classes of
vertices, the organization of the system and the function of its parts.

The community structure of complex networks
is still rather elusive. The definition of community is 
controversial, and should be adapted to the particular class
of systems/problems one considers. Consequently it is not yet clear
how scholars can test and validate community detection methods,
although the issue has lately received some
attention~\cite{lancichinetti08, lancichinetti09b,lancichinetti09c, orman10,orman11}. 
Also,  in order to deliver possibly more reliable results, 
methods should ideally exploit all features of the system, like edge
directedness and weight (for directed and weighted networks,
respectively), and account for properties of the partitions, like
hierarchy~\cite{sales07,clauset08} and community
overlaps~\cite{baumes05,palla05}. Very few methods are capable to take
all these factors into consideration~\cite{rosvall08,lancichinetti11}.
Another important barrier is the computational complexity of the
algorithms, which keep many of them from being applied to networks
with millions of vertices or larger.

In this paper we focus on another major problem affecting 
clustering techniques. Most of them, in fact, do not deliver a
unique answer. The most typical scenario is when the seeked
partition or individual clusters
correspond to extrema of a cost function~\cite{clauset05,newman06,lancichinetti09}, whose
search can only be carried out with approximation techniques, with
results depending on random seeds and on the choice of initial conditions. 
Allegedly deterministic methods may also run into similar
difficulties. For instance, in divisive clustering
methods~\cite{girvan02,radicchi04} the edges to be removed are the
ones corresponding to the lowest/highest value of a variable, and
there is a non-negligible chance of ties, especially in the final
stages of the calculation, when many edges
have been removed from the system. In such cases one usually picks at
random from the set of edges with equal (extremal) values, introducing
a dependence on random seeds. 

In the presence of several outputs of a given method, is there a partition
more representative of the actual community structure of the
system? If this were the case, one would need a criterion to sort out a
specific partition and discard all others. A better option is
combining the information of the different outputs into a new
partition. Exploiting the
information of different partitions is also very important in the
detection of communities in dynamic systems~\cite{hopcroft04, palla07,mucha10,chakrabarti06},
a problem of growing importance, given the increasing availability of
time-stamped network datasets~\cite{holme12}. 
Existing methods typically rely on the analysis
of individual snapshots, while the history of the system should also
play a role~\cite{chakrabarti06}. Therefore, combining partitions
corresponding to different time windows is a promising approach.

Consensus clustering~\cite{strehl02,topchy05,goder08} is a well known
technique used in data analysis to solve this problem. Typically, the goal is
searching for the so-called {\it median} (or {\it consensus}) {\it partition}, i.e. the partition
that is most similar, on average, to all the input partitions. The
similarity can be measured in several ways, for instance with the
Normalized Mutual Information (NMI)~\cite{danon05}. In its standard
formulation it is a difficult combinatorial
optimization problem. An alternative greedy strategy~\cite{strehl02}, which we
explore here, uses the {\it consensus
  matrix}, i.e. a matrix based on
the cooccurrence of vertices in clusters of the input
partitions. The consensus matrix is used as an input for the
graph clustering technique adopted, leading to a new set of
partitions, which generate a new consensus matrix, etc., until 
a unique partition is finally reached, which cannot be altered by
further iterations. This
procedure has proven to lead quickly to consistent and stable
partitions in real networks~\cite{kwak11}.

We stress that our goal is not finding a better optimum for the
objective function of a given method. Consensus partitions
usually do not deliver improved optima. On the other hand, global
quality functions, like modularity~\cite{newman04b}, are known to have
serious limits~\cite{fortunato07,good10,lancichinetti11b}, and
their optimization is often unable to detect clusters in realistic settings, not
even when the clusters are loosely connected to each other. In this
respect, insisting in finding the absolute optimum of the measure
would not be productive. However, if we buy the popular notion of
communities as subgraphs with a high internal edge density and a
comparatively low external edge density, the task of any method would
be easier if we managed to further increase the internal edge density
of the subgraphs, enhancing their cohesion, and to further decrease
the edge density between the subgraphs, enhancing their
separation. Ideally, if we could push this process to the extreme, we
would end up with a set of disconnected cliques, which every method
would be able to identify, despite its limitations. Consensus clustering
induces this type of transformation (Fig. 1) and therefore it mitigates the deficiencies  
of clustering algorithms, leading to more efficient techniques. The situation
in a sense recalls spectral clustering~\cite{luxburg06}, where by
mapping the original network in a network of points in a Euclidean
space, through the eigenvector components of a given matrix (typically
the Laplacian), one ends up with a system which is easier to clusterize.

In this paper we present the first systematic study of consensus clustering.
We show that the consensus partition gets much closer to the
actual community structure of the system than the partitions obtained
from the direct application of the chosen clustering method. 
We will also see how to monitor the evolution of clusters in temporal
networks, by deriving the consensus partition from several snapshots
of the system. We demonstrate the power of this approach by studying
the evolution of topics in the citation network of papers published by
the American Physical Society (APS).
\begin{figure}[h!]
\begin{center}
\includegraphics[width=\columnwidth]{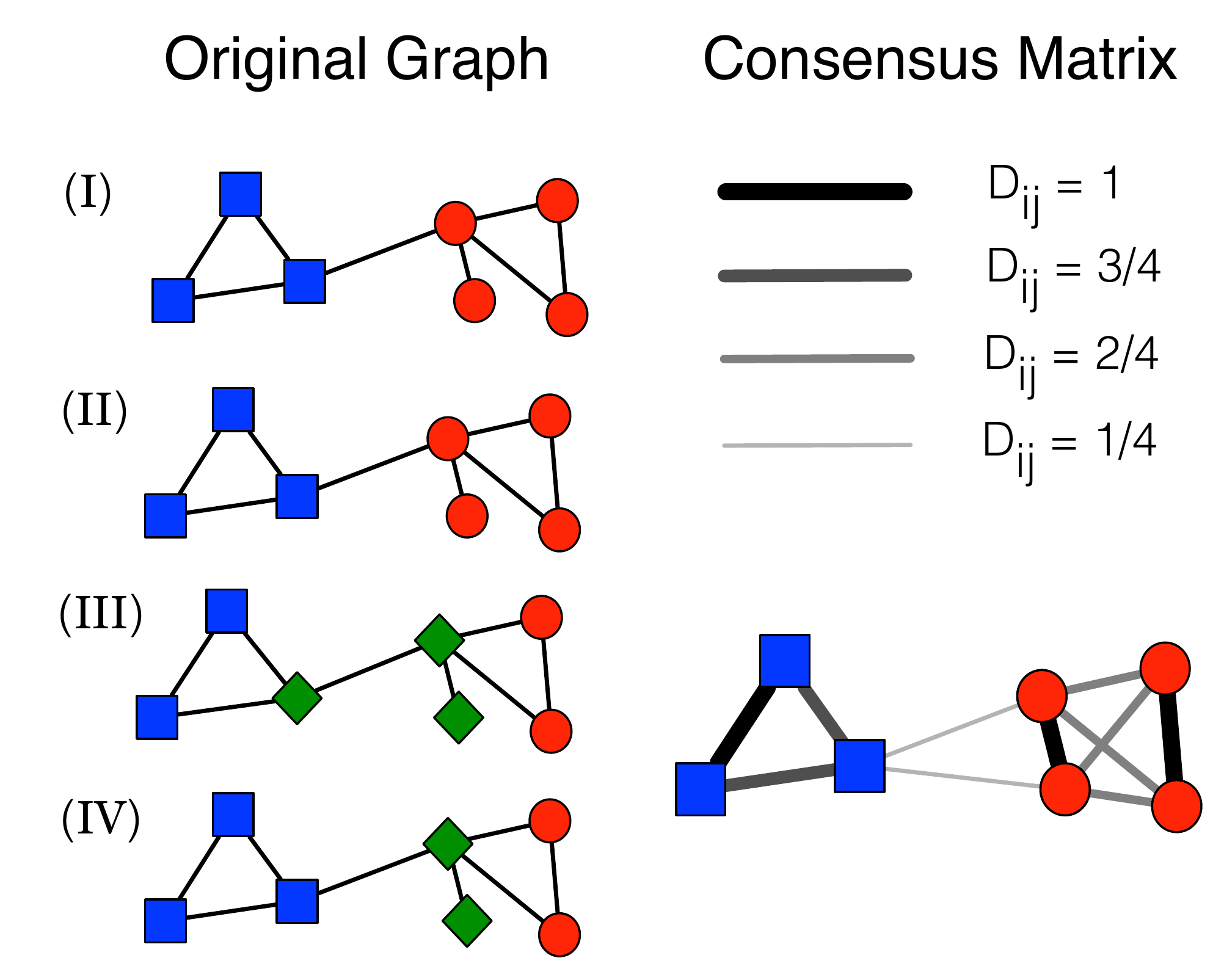}
\caption{Effect of consensus clustering on community structure.
Schematic illustration of consensus clustering on a graph with two
visible clusters, whose vertices are indicated by the squares and circles on the
(I) and (II) diagrams.
The combination of the partitions (I), (II), (III) and (IV) yields the
(weighted) consensus matrix illustrated on the right (see Methods). The thickness of
each edge is proportional to its weight. 
In the consensus matrix the cluster structure of the original network
is more visible: the two communities have become cliques, with
``heavy'' edges, whereas the connections between them are quite weak. Interestingly,
this improvement has been achieved despite the presence of two inaccurate
partitions in three clusters (III and IV).}
\label{fig1}
\end{center}
\end{figure}

\section{Results}

\subsection{Accuracy}

In order to demonstrate the superior performance achievable by
integrating consensus clustering in a given method, we tested the
results on artificial benchmark graphs with built-in community
structure. We chose the LFR benchmark graphs, which have become a
standard in the evaluation of the performance of clustering algorithms~\cite{lancichinetti08, lancichinetti09b,lancichinetti09c, orman10,orman11}.
The LFR benchmark is a generalization of the four-groups benchmark
proposed by Girvan and Newman, which is a particular realization of
the planted $\ell$-partition model by Condon and
Karp~\cite{condon01}. LFR graphs are characterized by power law
distributions of vertex degree and community size, features that
frequently occur in real world networks. 

The clustering algorithms we used are listed below:
\begin{itemize}
\item{{\it Fast greedy modularity optimization}. It is a technique developed
  by Clauset et al.~\cite{clauset04}, that performs a quick maximization of the
  modularity by Newman and Girvan~\cite{newman04b}. The accuracy of
  the estimate for the modularity maximum is not very high, but the
  method has been frequently used because it has been one of the first
techniques able to analyze large networks. We label it here as {\it Clauset
et al.}.}
\item{{\it Modularity optimization via simulated annealing}. Here the
    maximization of modularity is carried out in a more exhaustive
    (and computationally expensive) way. Simulated annealing is a
    traditional technique used in global optimization
    problems~\cite{kirkpatrick83}. The first application to modularity
  has been devised by Guimer\'a et al.~\cite{guimera04}. In contrast
  to the standard design, we
start at zero temperature. This is necessary because if the
method is very stable there is no point in using the consensus approach:  
if the algorithm systematically finds the same clusters, the consensus matrix $D$ would consist of 
$m$ disconnected cliques and the successive clusterization of $D$
would yield the same clusters over and over.
For the method we use the label {\it SA}.}
\item{{\it Louvain method}. The goal is still the optimization of
    modularity, by means of a hierarchical approach. First one
    partitions the original network in small communities, such
    to maximize modularity with respect to local moves of the
    vertices. This first generation clusters turn into supervertices
    of a (much) smaller weighted graph, where the procedure is
    iterated, and so on, until modularity reaches a maximum. It is a
    fast method, suitable to analyze very large graphs. However, like
    all methods based on modularity optimization, including the
    previous two, it is biased by the intrinsic limits of modularity
    maximization~\cite{fortunato07,good10, lancichinetti11b}. We refer
  to this method as to {\it Louvain}.} 
\item{{\it Label propagation method}. This method~\cite{raghavan07} 
    simulates the spreading of labels based on the simple rule that at each iteration a given vertex takes the
most frequent label in its neighborhood. The starting configuration is chosen such that every vertex is given a
different label and the procedure is iterated until convergence. This method has the problem of
partitioning the network such that there are very big clusters, due to the possibility
of a few labels to propagate over large portions of the graph. 
We considered asynchronous updates, i.e. we update the vertex
memberships according to the latest 
memberships of the neighbors. We shall refer to this method as {\it
  LPM}.}
\item{{\it Infomap}. The idea behind this method is the same as in
cartography: dividing the network in areas, like counties/states in
a map, and recycling the identifiers/names of vertices/towns among different
areas. The goal is to minimize the description of an
infinitely long random walk taking place on the
network~\cite{rosvall08}. When the graph has recognizable clusters, most of the
time the walker will be trapped within a cluster. That way, the additional cost of
introducing new labels to identify the clusters is compensated by the
fact that such labels are seldom used to describe the process, as transitions
between clusters are unfrequent, so the recycling of the binary identifiers
for the vertices among different clusters leads to major savings in the description of the
random walk. We shall refer to this method as {\it
  Infomap}.}
\item{{\it OSLOM}. The method relies on the concept of statistical
    significance of clusters. The idea here is that, since random
    graphs are not supposed to have clusters, the subgraphs of a
    network that are deemed to be communities should be very different
    from the subgraphs one observes in a random graph with similar
    features as the system at study. The statistical significance is
    then estimated through the probability of finding the observed clusters in a
random network with identical expected degree
sequence~\cite{lancichinetti11}. Clusters are identified by
maximizing locally such probability.
We shall refer to this method as {\it
  OSLOM}.}
\end{itemize}

All the above techniques can be applied to weighted networks, a necessary
requisite for our implementation of consensus clustering (see
Methods).
\begin{figure*}[htb]
\begin{center}
\includegraphics[width=\textwidth]{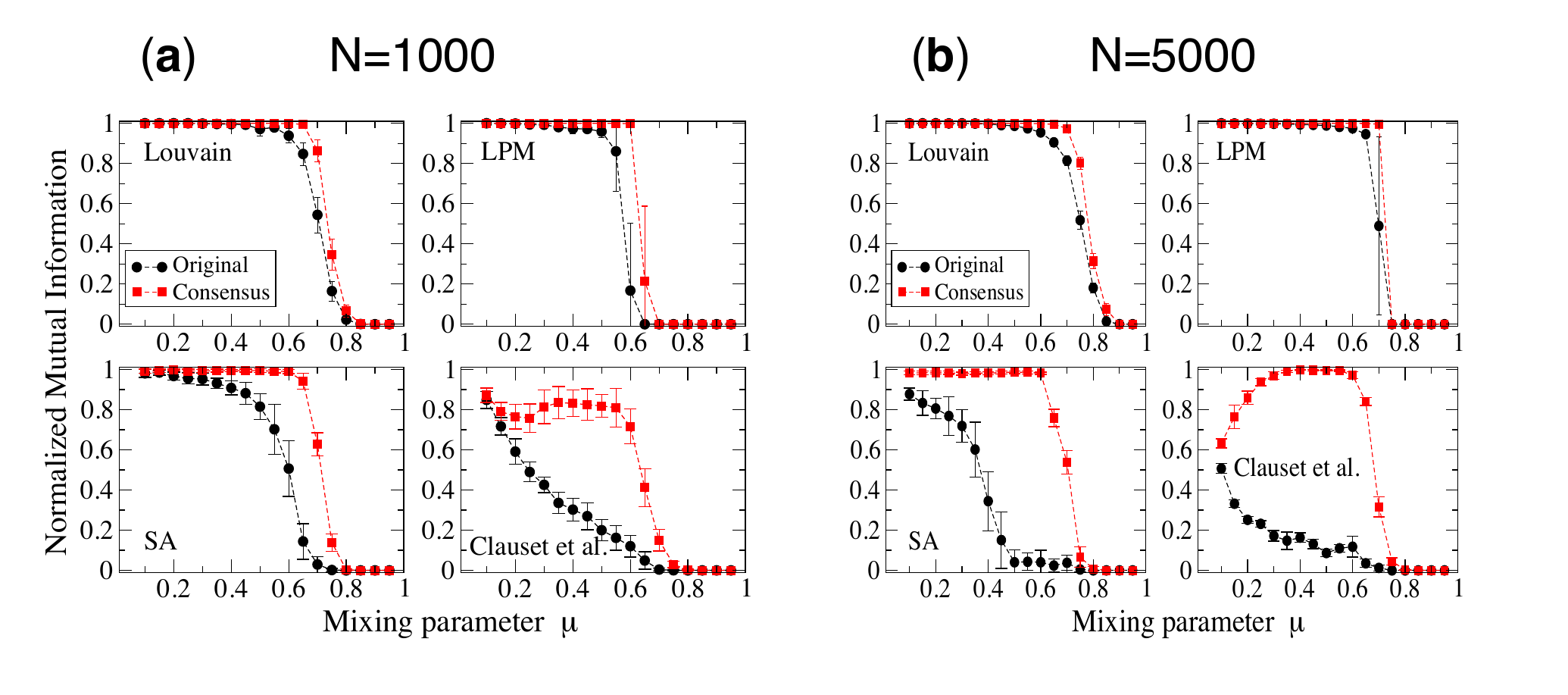}
\caption{Consensus clustering on the LFR benchmark. The dots indicate
the performance of the original method, the squares that obtained with
consensus clustering. The parameters of the LFR benchmark graphs are: 
average degree $\langle k \rangle=20$, maximum degree $k_{max}=50$,
minimum community size $c_{min}=10$, maximum community size
$c_{max}=50$, the degree exponent is $\tau_1=2$, the community size
exponent is $\tau_2=3$. Each panel correspond to a clustering
algorithm, indicated by the label. The two sets of plots correspond to networks with $1000$
(a) and $5000$ (b) vertices.}
\label{fig2}
\end{center}
\end{figure*}

In Fig. 2 we show the results of our tests. Each panel reports the
value of the Normalized Mutual Information (NMI) between the planted
partition of the benchmark and the one found by the algorithm as a
function of the mixing parameter $\mu$, which is a measure of the
degree of fuzziness of the clusters. Low values of $\mu$ correspond to
well-separated clusters, which are fairly easy to detect; by
increasing $\mu$ communities get more mixed and clustering
algorithms have more difficulties to distinguish them from each
other. As a consequence, all curves display a decreasing trend. The
NMI equals $1$ if the two partitions to compare are identical, and
approaches $0$ if they are very different. In Fig 2a and 2b the
benchmark graphs consist of $1000$ and $5000$ vertices,
respectively. Each point corresponds to an average over $100$
different graph realizations. For every realization we have produced
$150$ partitions with the chosen algorithm. The curve ``Original''
shows the average of the NMI between each partition and the planted
partition. The curve ``Consensus'' reports the NMI between the
consensus and the planted partition, where the former has been derived
from the $150$ input partitions. We do not show the results for Infomap and OSLOM because their performance on the LFR
benchmark graphs is very good
already~\cite{lancichinetti09c,lancichinetti11}, so it could not be
sensibly improved by means of consensus clustering (we have
verified that there still is a small improvement, though). 
The procedures to set the number of runs and the value of the
threshold $\tau$ for each method are detailed in the Appendix.
In all cases, consensus clustering leads to better partitions than
those of the original method. The improvement is particularly
impressive for the method by Clauset et al.: the latter is known to have a poor
performance on the LFR benchmark~\cite{lancichinetti09c}, and yet in
an intermediate range of
values of the mixing parameter $\mu$ it is able to detect the right partition
by composing the results of individual runs. For $\mu$ small the
algorithm delivers rather stable results, so the consensus partition
still differs significantly from the planted partition of the
benchmark.
In the Appendix we give a mathematical argument
to show why consensus clustering is so effective on the LFR benchmark.

\subsection{Stability}

Another major advantage of consensus clustering is the fact that it
leads to stable partitions~\cite{kwak11}. Here we verify how stability
varies with the number of input runs $r$. 
\begin{figure}[h!]
\begin{center}
\includegraphics[width=\columnwidth]{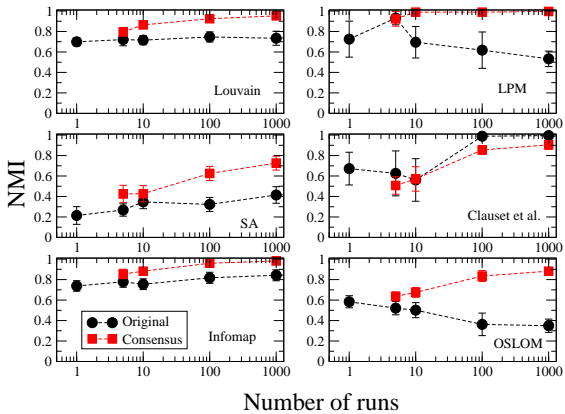}
\caption{Stability plot for the neural network of {\it
    C. elegans}. The network has $453$ vertices and $2\,050$
edges.}
\label{fig3}
\end{center}
\end{figure}
\begin{figure}[h!]
\begin{center}
\includegraphics[width=\columnwidth]{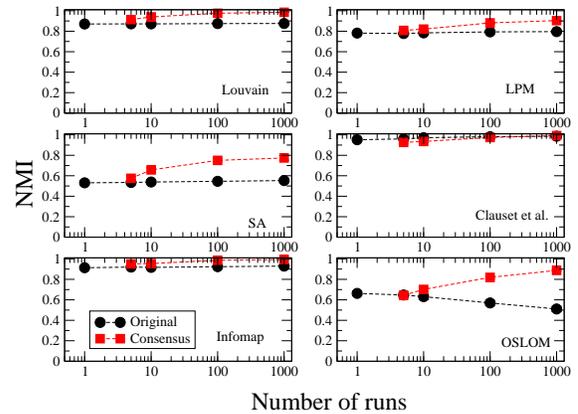}
\caption{Stability plot for the citation network of papers published in
journals of the American Physical Society (APS). The original dataset
is too large to get results in a reasonable time, so the plot refers
to the subset containing all papers published in $1960$ and the ones
cited by them.}
\label{fig4}
\end{center}
\end{figure}
In Figs. 3 and 4 we present
stability plots for two real world datasets: the neural network of
{\it C. elegans}\cite{white86,watts98} ($453$ vertices, $2\,050$
edges); the citation network of papers published in
journals of the American Physical Society (APS) ($445\,443$
vertices, $4\,505\,730$ directed edges). Each figure shows two curves:
the average NMI between best partitions (circles);
the average NMI between consensus
partitions (squares). Both the best and the consensus partition are
computed for $r$ input runs, and the procedure is repeated for $20$
sequences of $r$ runs. So we end up having $20$ best partitions and
$20$ consensus partitions. The values
reported are then averages over all possible pairs that one can have
out of $20$ numbers. 
Each of the six panels corresponds to a
specific clustering algorithm. To derive
the consensus partitions we used the
same values of the threshold parameter $\tau$ 
as in the tests of Fig. 2a
(for Infomap and OSLOM $\tau=0.5$).

As ``best'' partition for
Louvain, SA and Clauset et al. we take the one with largest
modularity. This sounds like the most natural choice, since such methods aim
at maximizing modularity. For the LPM there is no way to determine
which partition could be considered the best, so we took the one with
maximal modularity as well. On the other hand, both Infomap and OSLOM have
the option to select the best partition out of a set of $r$ runs. 

In Fig. 3 we show the stability plot for {\it C. elegans}. For all
methods the consensus partition turns out to be more stable of
the input than the best partition. The only exception is the method by
Clauset et al., but the two curves are rather close to each other. We
remark that increasing the number of input runs does not necessarily
imply more stable partitions. In the cases of LPM and OSLOM, for instance,
the best partitions of the method get more unstable for $r\simeq 10$. On
the other hand, the stability of the consensus partition is
monotonically increasing for all six algorithms. 

In Fig. 4 we see the corresponding plot for the APS dataset. The
analysis of the full dataset is too computationally expensive, so we
focused on a subset, that of papers published in $1960$, along with
the papers cited by them. The
resulting network has $5\,696$ vertices and $8\,634$ edges. 
Again, we see that the stability of the consensus partition grows
monotonically with the number of input runs $r$, and it remains
higher than that of the best partition.

In the Appendix
we show that the consensus partition is not only more stable,
but it also has higher fidelity than the individual input partitions it
combines (Figs. S5 and S6).

\subsection{Dynamic communities}

Consensus clustering is a powerful tool to explore
the dynamics of community structure as well. Here we show that it is able
to monitor the history of the citation network of the APS, and to
follow birth, growth, fragmentation, decay and death of scientific
topics. The procedure to derive the consensus partitions out of time
snapshots of a network is described in the Methods.
\begin{figure*}[htb]
\begin{center}
\includegraphics[width=\textwidth]{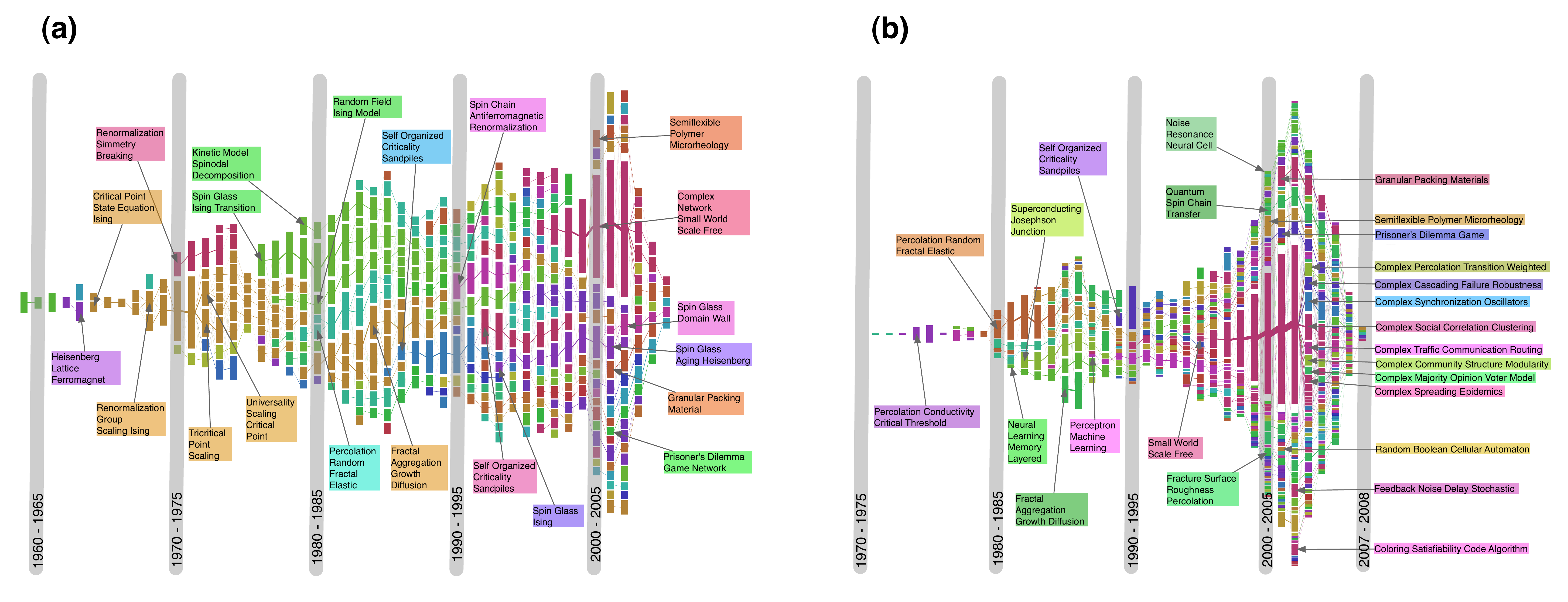}
\caption{Time evolution of clusters in the APS citation
network. In (a) we selected all the clusters that have at least 
one of the keywords {\it Criticality}, {\it Fractal}, {\it Ising},  {\it Network} and {\it  Renormalization}
among the top 15 most frequent words appearing in the title of the
papers, while in (b) we just filtered the keyword {\it Network(s)}. Both diagrams were obtained using Infomap on 
snapshots spanning each a window of $5$ years, except at the right end
of each diagram: since there is no data after $2008$, the last windows
must have $2008$ as upper limit, so their size shrinks ($2004-2008$, $2005-2008$,
$2006-2008$, $2007-2008$). Consensus is computed by
combining pairs of consecutive snapshots (see Methods). A color
uniquely identifies a module, while
the width of the links between clusters is proportional to the number of papers they have in
common. In (b) we observe the rapid growth of the field {\it
  Complex Networks}, which eventually splits in a number of smaller
subtopics, like {\it Community Structure}, {\it
  Epidemic Spreading}, {\it Robustness}, etc.. }
\label{fig5}
\end{center}
\end{figure*}
The evolution of the APS dataset is shown in Fig. 5a. The system is too
large to be meaningfully displayed in a single figure, so we focused on the evolution of communities of papers in
Statistical Physics. For that, we selected only the clusters whose papers include 
{\it Criticality}, {\it Fractal}, {\it Ising},  {\it Network}  and {\it
  Renormalization} among the $15$ most frequent words
in their titles. Each vertical bar corresponds to a time window of $5$
years (see Methods), its length
to the size of the system. The time ranges from
$1945$ until $2008$. The evolution is characterized by alternating
phases of expansion and contraction, although in the long term there is a growing
tendency in the number of papers. This is due to the fact that 
the keywords we selected were fashionable in different historical
phases of the development of Statistical Physics, so some of them
became obsolete after some time (i.e., there are less papers with
those keywords), while at the same time others become more fashionable. 
Communities are identified by the colors. 
Pairs of matching clusters in consecutive times are marked by the same
color. Clusters of consecutive time windows sharing papers are joined
by links, whose width is proportional to the number of common papers.
We mark the clusters corresponding to famous topics in Statistical
Physics, indicating the most frequent words appearing in the
titles of the papers of each cluster.
One can spot the emergence of new fields, like {\it Self-Organized
Criticality}, {\it Spin Glasses} and {\it Complex Networks}. 

In Fig. 5b we consider only papers with the words {\it Network} or
{\it Networks} among the $15$ most frequent words in their
titles. Here we can observe the genesis of the fields {\it Neural
  Networks} and {\it Complex Networks}. In order to have clearer
pictures, in Fig. 5a we only plotted clusters that have at least 50
papers, while in Fig. 5b the threshold is 10 papers. 
\begin{figure}[h!]
\begin{center}
\includegraphics[width=\columnwidth]{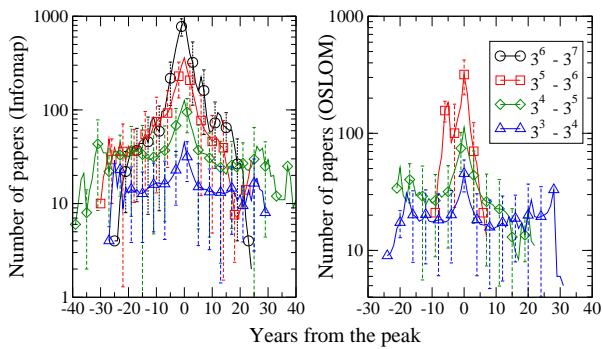}
\caption{Evolution of average size of clusters. The
time ranges of the evolution of  the communities have been shifted
such that the year when a cluster reaches its maximum is $0$. The two
panels show the results obtained with Infomap (left) and OSLOM
(right). The data are aggregated in four bins, according to the 
maximum size reached by the cluster. The phases of growth and decay of
fields appear rather symmetric.}
\label{fig6}
\end{center}
\end{figure}
For a quantitative assessment of the birth, evolution and death of topics, we keep track of
each cluster matching it with the most similar module in the following
time frame (see Methods). This allows us to
compute one sequence for each cluster, which reports its size for all the years when the community was
present. In Fig. 6 we computed the statistics of these sequences, centering them on the year when the cluster
reached its peak (reference year $0$). To obtain smooth patterns,
clusters are aggregated in bins according to their peak magnitude. 
Fig. 6 shows the average cluster size for each bin as a function of the years from the
peak. We computed the curves using Infomap (left) and OSLOM (right). Around the peak, the
cluster sizes are highly heterogeneous, with some important topics reaching almost
$1000$ papers at the peak (for Infomap). The rise and decline of
topics take place around 10 years before and after the peak,
with a remarkably symmetric pattern with respect to the maximum.

\section{Discussion}

Consensus clustering is an invaluable tool to cope
with the stochastic fluctuations in the results of clustering
techniques. We have seen that the integration of consensus clustering
with popular existing techniques leads to more accurate partitions
than the ones delivered by the methods alone, in artificial graphs
with planted community structure. This holds even for methods whose
direct application gives poor results on the same graphs. In this way it is possible to fully
exploit the power of each method and the diversity of the partitions,
rather than being a problem, becomes a factor of performance enhancement.

Finding a consensus between different partitions also offers a natural
solution to the problem of detecting communities in dynamic networks. 
Here one combines partitions corresponding to snapshots of the system,
in overlapping time windows. Results depend on the choice of the
amplitude of the time windows and on the number of snapshots combined
in the same consensus partition. The choice of these parameters may
be suggested by the specific system at study. It is usually possible
to identify a meaningful time scale for the evolution of the
system. In those cases both the size of the time windows and the
number of snapshots to combine can be selected accordingly. As a safe
guideline one should avoid merging partitions referring to a time
range which is much broader than the natural time scale of the network.
A good policy is to
explore various possibilities and see if results are robust within
ample ranges of reasonable values for the parameters. Additional
complications arise from the fact that the evolution of the system may
not be linear in time, so that it cannot be followed in terms of
standard time units. In citation networks, like the one we studied,
it is known that the number of published papers has been increasing
exponentially in time. Therefore, a fixed time window would cover many
more events (i.e. published papers and mutual citations) if it refers to a recent period than to some decades ago. 
In those cases, a natural choice could be to consider snapshots
covering time windows of decreasing size.

\section*{Methods}
{\bf The consensus matrix}. Let us suppose that we wish to combine $n_P$
partitions found by a clustering algorithm on a network with $n$ vertices. The consensus matrix ${\bf
  D}$ is an $n\times n$ matrix, whose entry $D_{ij}$ indicates the
number of partitions in which vertices $i$ and $j$ of the network were
assigned to the same cluster, divided by the number of partitions
$n_P$. The matrix ${\bf D}$ is usually much denser than the adjacency
matrix ${\bf A}$ of the original network, because
in the consensus matrix there is an edge between any two vertices
which have cooccurred in the same cluster at least once. On the other
hand, the weights are large only for those vertices which are most
frequently co-clustered, whereas low weights indicate that the
vertices are probably at the boundary between different (real)
clusters, so their classification in the same cluster is unlikely and
essentially due to noise. We wish to maintain the large weights and to
drop the low ones, therefore a filtering procedure is in order. 
Among the other things, in the absence of filtering the consensus matrix
would quickly grow into a very dense matrix, which would make the application of
any clustering algorithm computationally expensive. 

We discard all entries of ${\bf D}$
below a threshold $\tau$. We stress that there might be some noisy
vertices whose edges could all be below the threshold,  
and they would be not connected anymore. When this happens, we just
connect them to their neighbors with highest weights, to keep the
graph connected all along the procedure.

Next we apply the same clustering algorithm to ${\bf D}$ and produce
another set of partitions, which is then used to construct a new
consensus matrix ${\bf D}^{\prime}$, as described above. The procedure
is iterated until the consensus matrix turns into a block diagonal
matrix ${\bf D}^{final}$, whose weights equal $1$ for vertices in the same block and $0$
for vertices in different blocks. The matrix ${\bf D}^{final}$
delivers the community structure of the original network. In our
calculations typically one iteration is sufficient to lead to stable
results. We remark
that in order to use the same clustering method all along,
the latter has to be able to detect clusters in weighted networks,
since the consensus matrix is weighted. 
This is a necessary constraint on the choice of the methods for which
one could use the procedure proposed here. However, it is not a severe
limitation, as most clustering algorithms in the literature can handle
weighted networks or can be trivially extended to deal with them.

We close by summarizing the procedure, step by step. The starting
point is a network ${\cal G}$ with $n$ vertices and a clustering
algorithm ${\bf A}$.
\begin{enumerate}
\item{Apply ${\bf A}$ on ${\cal G}$ $n_P$ times, so to yield $n_P$ partitions.}
\item{Compute the consensus matrix ${\bf D}$, where $D_{ij}$ is the
number of partitions in which vertices $i$ and $j$ of ${\cal G}$ are
assigned to the same cluster, divided by $n_P$.}
\item{All entries of  ${\bf D}$
below a chosen threshold $\tau$ are set to zero.}
\item{Apply ${\bf A}$ on ${\bf D}$ $n_P$ times, so to yield $n_P$ partitions.}
\item{If the partitions are all equal, stop (the consensus matrix
    would be block-diagonal). Otherwise go back to 2.}
\end{enumerate}

\vskip0.5cm
{\bf Consensus for dynamic clusters}. In the case of temporal
networks, the dynamics of the system is represented as a succession of
snapshots, corresponding to overlapping time windows. Let us suppose
to have $m$ windows of size $\Delta t$ for a time range going from $t_0$ to $t_m$. 
We separate them as $[t_0, t_0+\Delta t]$, $[t_0+1, t_0+\Delta t+1]$,
$[t_0+2, t_0+\Delta t+2]$, ...,
$[t_{m}-\Delta t, t_m]$. Each time window is shifted by one time unit
to the right with respect to the previous one.
The idea is to derive the consensus partition from subsets of $r$ consecutive
snapshots, with $r$ suitably chosen. One starts by combining the first
$r$ snapshots, then those from $2$ to $r+1$, and so on until the 
interval spanned by the last $r$ snapshots. In our calculations for
the APS citation network we
took $\Delta t=5$ (years), $r=2$.

There are two sources of fluctuations: 1) the ones coming from the different
partitions delivered by the chosen clustering technique for a given
snapshot; 2) the ones coming from the fact that the structure of the
network is changing in time.
The entries of the consensus matrix $D_{ij}$ are obtained by computing the
number of times vertices $i$ and $j$ are clustered together, and
dividing it by the number of partitions corresponding to snapshots
including both vertices. This looks like a more sensible choice with respect
to the one we had adopted in the static case (when we took the total
number of partitions used as input for the consensus matrix), as in the evolution of a temporal network
new vertices may join the system and old ones may disappear. 

Once the consensus partitions for each time step have been derived, 
there is the problem of relating clusters at different times. We need
a quantitative criterion to establish whether a cluster ${\cal
  C}_{t+1}$ at time $t+1$ is the evolution of a cluster ${\cal
  C}_{t}$ at time $t$. The correspondence is not trivial: a cluster may fragment, and thus there would be many
``children'' clusters at time $t+1$ for the same cluster at time
$t$. In order to assign to each cluster ${\cal
  C}_{t}$ of the consensus partition at time $t$ one and only one
cluster of the consensus partition ${\cal P}^{t+1}$ at time $t+1$ we compute the Jaccard
index~\cite{jaccard01} between ${\cal C}_{t}$ and every cluster of ${\cal P}^{t+1}$, and
pick the one which yields the largest value. The Jaccard index
$J(A,B)$ between two sets $A$ and $B$ equals
\begin{equation}
J(A,B)=\frac{|A\cap B|}{|A\cup B|}.
\end{equation}
In our case, since the snapshots generating the partitions refer to
different moments of the life of the system and may not
contain the same elements, the Jaccard index is computed by excluding
from either cluster the vertices which are not present in both
partitions. The same procedure is followed to assign to each cluster 
${\cal C}_{t+1}$ of the consensus partition at time $t+1$ one and only one
cluster of the consensus partition ${\cal P}^{t}$ at time $t$.  In
general, if cluster $A$ at time $t$ is the best match of cluster $B$
at time $t+1$, the latter may not be the best match of $A$. If it is,
then we use the same color for both clusters. Otherwise there is a
discontinuity in the evolution of $A$, which stops at $t$, and its
best match at time $t+1$ will be considered as a newly born cluster.

\appendix

\section{Selection of the optimal number of runs and threshold $\tau$}

For any implementation of our method there are two parameters that
need to be set before starting the computation: 1) the number $r$ of
partitions to be combined in the consensus matrix (see Methods); 2)
the threshold $\tau$ used to filter the entries of the consensus
matrix, to avoid that the latter becomes too dense, slowing down the procedure.
In Figs. 7 and 8 we show how these numbers are chosen. 
Fig. 7 displays the Normalized Mutual Information (NMI) between the
consensus partition and the planted partition of the benchmark graphs
used for Fig. 1, for different values of $r$ and a specific value of
the mixing parameter $\mu$. Each curve corresponds
to a different value for the threshold $\tau$, which equals $0$, $0.5$
and $0.7$. Each panel presents the result of a different clustering
algorithm; the value of $\mu$ varies for each method because consensus
is the most effective the more diverse the input partitions
are. Therefore we picked the value of $\mu$ at which
the original method starts failing ($\mu=0.7$ for Louvain, $\mu=0.6$
for the LPM, $\mu=0.65$ for SA and $\mu=0.4$ for Clauset et
al.. From Fig. 7 we deduce that for $r \approx 50$ one reaches an optimal
partition with consensus clustering, which remains stable for larger
values. This seems to hold regardless of the value of the threshold
parameter $\tau$.
Therefore, in our tests of Fig. 1 we have taken $r$ between $50$ and $100$. 
\begin{figure}
\begin{center}
\label{figS1}
\includegraphics[width=\columnwidth]{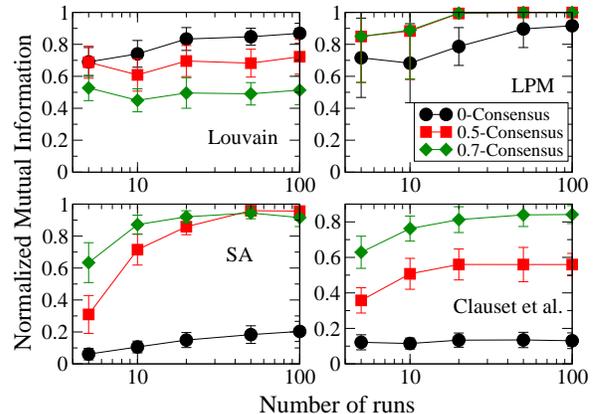}
\caption{Accuracy of consensus clustering with the number of input
  runs $r$. Each panel reports the NMI between the planted partition
  of the LFR benchmark graphs used for Fig.~1, at a given $\mu$ (see text), as a
  function of the number of runs for a specific method. The symbols
  refer to three different choices for the threshold parameter $\tau$:
  $0$ (circles), $0.5$ (squares), $0.7$ (diamonds).}
\end{center}
\end{figure}
\begin{figure}
\begin{center}
\label{figS2}
\includegraphics[width=\columnwidth]{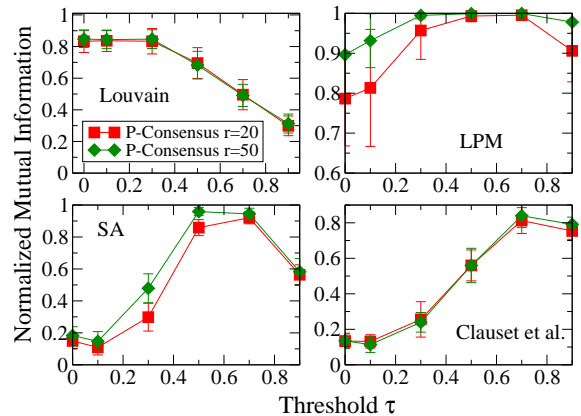}
\caption{Accuracy of consensus clustering with the threshold parameter
  $\tau$. Each panel reports the NMI between the planted partition
  of the LFR benchmark graphs used for Fig.~7, as a
  function of $\tau$ for a specific method. The symbols
  refer to two different choices for the number of runs $r$:
  $20$ (squares), $50$ (diamonds).}
\end{center}
\end{figure}

In Fig. 8 we show instead how the threshold $\tau$ affects the
results. We use the same benchmark graphs as in Fig. 7, and two
values for the number of runs $r$: $20$ and $50$. The y-axis reports
again the value of the NMI between the consensus and the planted
partition of the benchmark. We see that the ranges of optimal values
for $\tau$ depend on the clustering technique adopted. For Louvain, it
is best to choose a low threshold, for the LPM $\tau$-values in the
range $[0.3, 0.7]$ give optimal results, for SA the best value span a
shorter range (from $0.5$ to $0.7$) and for Clauset et al. the best
results are obtained for fairly high values of the threshold.

\section{On the effectiveness of consensus clustering}

To understand why consensus clustering is so effective at detecting
the clusters of the LFR benchmarks, we discuss here a simpler model
which resembles the LFR benchmarks. 
We focus on modularity optimization because it is easier to understand how consensus improves the method.

We consider a graph with $C$ cliques
of $n_c$ vertices each. The cliques are connected by placing $C*h$ edges between randomly
chosen pairs of vertices, where the vertices of each pair belong to 
different cliques. It can be proven that for this kind of graph, the modularity function
is optimal for a partition of $\sqrt{M}$ modules of equal size, where $M$ is the number of edges. This
result has been proved for a ring of cliques, but it is straightforward to verify that the same proof can be extended to this case.

Since there is a high number of combinations to group the cliques
together in order to reach the optimal number 
of modules, we expect that, on average, each clique will be joined to
some of its neighboring cliques with roughly 
equal probability. If we call $g$ the average number of neighboring
cliques, the probability that two neighboring cliques are found in the same
cluster is simply $\frac{g}{2h}$ (we recall that we placed $C*h$ links).

If $C \gg 1$ and $h$ is small enough so that the network of cliques is
sparse, there will be a very small number of edges 
between the cliques grouped in the same module. The smallest number of
edges necessary to keep $n$ vertices connected 
is $n-1$ (which gives a tree-like structure), and in such a case every vertex has an
average degree $ \langle k \rangle  \approx 2$, 
when $n \gg 1$. In the case of a tree, we would have that $g \approx 2$.
The probability for two cliques to be connected if their distance in the clique network is $d$, will be, in general,

\begin{equation}
p_d \approx \frac{1}{h (2h-1)^{d-1}}.
\end{equation}

Fig.~9 shows that this approximation is not bad especially for high values of $C$.
In the following plots we considered $h=10$, $n_c=10$.

\begin{figure}[h!]
\begin{center}
\includegraphics[width=\columnwidth]{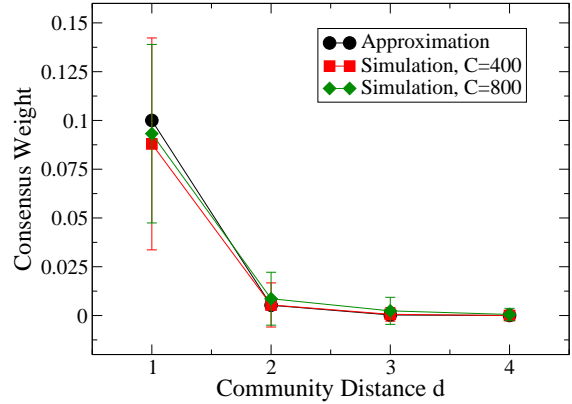}
\caption{Probability of joining cliques in our stylized model graph
  with communities. Consensus weight indicates the probability that a pair
  of cliques are joined together. This is plotted as a function of the
  distance between cliques in the community network, i.e. the graph where cliques are
  to be seen as supervertices.}
\end{center}
\end{figure}

Eventually, we might expect that choosing a value of the threshold $\tau > p_1$, the consensus matrix will consist of $C$ disconnected cliques.
Modularity optimization
instead would always merge cliques together in larger clusters.
Indeed, the NMI between the planted partition and the input partitions
is much lower than the NMI between the planted and the consensus
partition already for $\tau = 0.3$ (Fig.~10). 

\begin{figure}[h!]
\begin{center}
\includegraphics[width=\columnwidth]{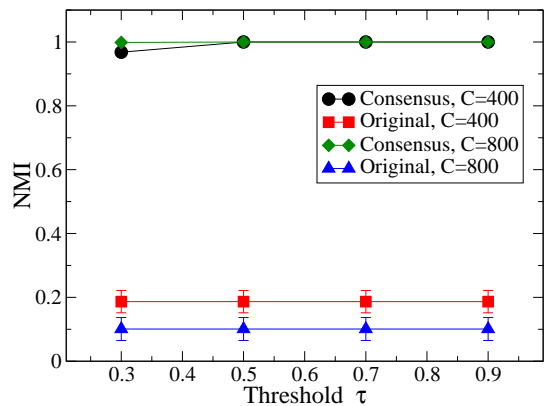}\caption{Fidelity of consensus partition. Normalized Mutual Information between the planted partition
  and the input partitions (squares and diamonds) and between the
  planted and the consensus partition (circles and triangles) as a
  function of the threshold $\tau$. We consider two values for the
  number of cliques of the model network, $400$ and $800$.}
\end{center}
\end{figure}

\section{Stability and fidelity of consensus partitions}

In Figs. 3  and 4 we have shown that the consensus
partitions are more stable than the best partitions.  
However trivial partitions, like the one where all vertices
are together in the same cluster, would be the most stable possible, 
although they would be completely unrelated to the input
partitions. To prove that the consensus partitions are 
actually very close to the input ones, Fig.~11 shows the Normalized
Mutual Information of the input partitions among themselves (the average value of NMI among all pairs
of different partitions) and the average value of NMI between the input partitions and the
consensus partition, for different values of the threshold $\tau$, for the neural network of
\textit{C. elegans}. Fig.~12 shows the same plot for the APS citation network of papers
published in 1960. Indeed the consensus partition is often even closer to the input
partitions than the latter are to each other, with the additional advantage of
being more stable. This does not hold only when the
threshold is too high, because in this case the consensus partition is
made of small clusters, since too many
connections carry a weight under the threshold and are deleted. Otherwise this should explain why the
consensus partition is more representative than the input. In Figs.~11 and 12
we considered $50$ input partitions, but the results are practically
the same if we take $100$ of them.
\begin{figure}[h!]
\begin{center}
\includegraphics[width=\columnwidth]{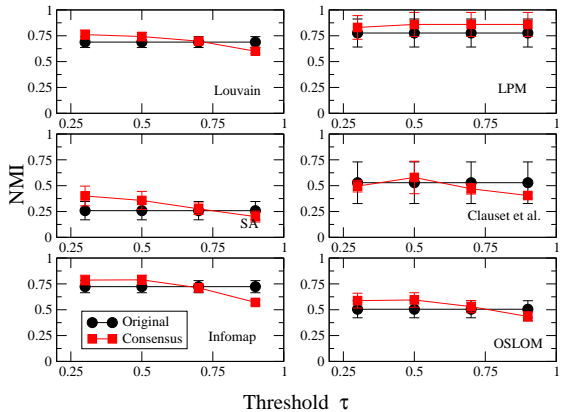}
\caption{Fidelity of consensus partition. The black curve reports the average NMI between pairs of 
input partitions, the red one is the average NMI between the 
input partitions and the consensus one, for the neural network of \textit{C. elegans}.}
\end{center}
\end{figure}
\begin{figure}[h!]
\begin{center}
\includegraphics[width=\columnwidth]{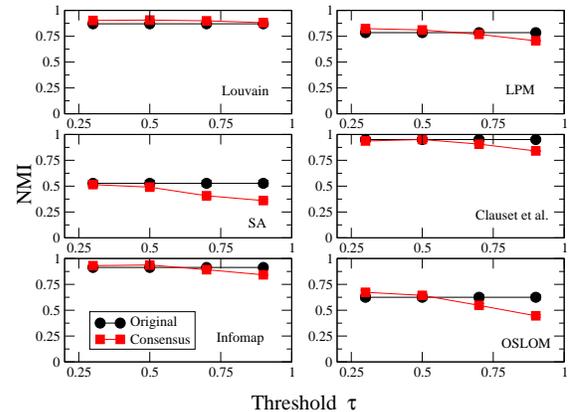}
\caption{Fidelity of consensus partition. Same plot as Fig.~11 for a snapshot of the APS citation network, with the papers published in 1960 and those cited by them.}
\end{center}
\end{figure}

\begin{acknowledgments}
We gratefully acknowledge ICTeCollective, grant 
238597 of the European Commission.
\end{acknowledgments}

\end{document}